\documentclass[prl,twocolumn,showpacs,preprintnumbers,amsmath,amssymb]{revtex4}

\usepackage{graphics}
\usepackage{graphicx}
\usepackage{dcolumn}
\usepackage{bm}

\usepackage{amssymb}
\usepackage{amsmath}

\font\tenscr=rsfs10 scaled1100
\font\sevenscr=rsfs7 
\font\fivescr=rsfs5 
\skewchar\tenscr='177 
\skewchar\sevenscr='177
\skewchar\fivescr='177 
\newfam\scrfam 
\textfont\scrfam=\tenscr
\scriptfont\scrfam=\sevenscr 
\scriptscriptfont\scrfam=\fivescr

\def\scri{{\fam\scrfam I}} 

\def\O{\mathcal{O}}

\begin{document}

\title{Early radiative properties of the developments of time
  symmetric, conformally flat initial data}

\author{Juan Antonio Valiente Kroon}
 \email{jav@aei-potsdam.mpg.de}
\affiliation{
Max Planck Institut f\"{u}r Gravitationsphysik, Albert Einstein
Institut, Am M\"{u}hlenberg 1, 14476 Golm, Germany. 
}

\date{\today}

\begin{abstract}
Using a representation of spatial infinity based in the properties of
conformal geodesics, the first terms of an expansion for the Bondi
mass for the development of time symmetric, conformally flat initial
data are calculated. As it is to be expected, the Bondi mass agrees
with the ADM at the sets where null infinity ``touches'' spatial
infinity. The second term in the expansion is proportional to the sum
of the squared norms of the Newman-Penrose constants of the spacetime.
In base of this result it is argued that these constants may provide a
measure of the incoming radiation contained in the spacetime. This is
illustrated by means of the Misner and Brill-Lindquist data sets.
\end{abstract}

\maketitle

\emph{Introduction.} The study of the radiative properties of
dynamical spacetimes describing isolated systems adquires these days
more relevance in view of the beginning of operations of several
interferometric gravitational wave detectors (LIGO, GEO 600). One of
the crucial radiative properties to be analysed is the mass loss due
to the radiative processes. Strictly speaking, the radiative
properties are global features of the spacetimes, and thus they can
only be unambiguosly defined at infinity. Penrose \cite{Pen63} has
put forward the idea of the description of spacetimes describing
isolated system by means of an \emph{unphysical spacetime}
$(M,g_{\mu\nu})$ obtained from the original (physical) spacetime
$(\widetilde{M},\widetilde{g}_{\mu\nu})$ through a conformal
rescaling: $g_{\mu\nu}=\Theta^2\widetilde{g}_{\mu\nu}$. It is within
this framework that notions like the mass loss due to radiative
processes can be rigorously defined. The mass loss
notion is captured in the form of the \emph{Bondi mass} which is
given as the integral \cite{PenRin86},
\begin{equation}
\label{bondi_mass}
m_B=-\frac{1}{4\pi}\oint
\left(\psi_2^0-\sigma_0\dot{\overline{\sigma}}_0 \right) dS,
\end{equation}
over cuts of null infinity. The functions $\psi_2^0$ and $\sigma_0$
are the leading terms in of the component $\widetilde{\psi_2}$
($\O(\Theta^3)$)of the Weyl tensor and the Newman-Penrose spin
coefficient $\widetilde{\sigma}$ ($\O(\Theta^2)$) respectively in a
so-called Bondi system.

To the author's knowledge, the formula (\ref{bondi_mass}) has never
been evaluated on a realistic spacetime. Furthermore, the dependence
of the Bondi mass on the initial data (assuming that the spacetime
arises as the development of some suitable Cauchy data)
whose development gives rise to (weakly) asymptotically simple
spacetimes has never been analysed. It is the purpose of this letter
to show that by means of the conformal field equations and an adequate
gauge choice it is possible
to obtain expansions of the gravitational field in which the
dependence on initial data can be read directly. In particular, this
approach enables us to calculate an expansion for the Bondi mass
which shows a remarkable, yet misterious, connection with another
global property of the spacetime: the Newman-Penrose constants ---see
equation (\ref{bondi_mass:expansion}). Our analysis will be
restricted to time symmetric, conformally flat initial data.    

\begin{figure}[t]
\includegraphics[width=.2\textwidth]{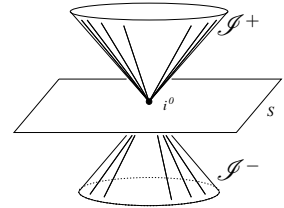}
\put(-25,65){$\scri^+$} 
\put(-25,10){$\scri^-$}
\caption{\label{fig:point} The standard representation of spacetime
  in the neighbourhood of spatial infinity. Spacetime corresponds to
  the region \emph{outside} the cone.}
\end{figure}
\begin{figure}[t]
\includegraphics[width=.25\textwidth]{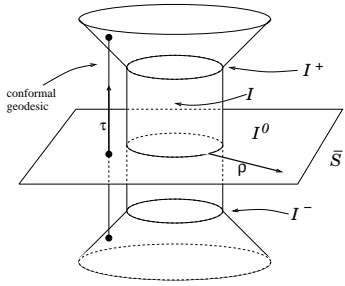} 
\caption{\label{fig:cylinder} The cylinder at spatial infinity. Again,
  spacetime corresponds to the region outside the cones and the
  cylinder.}
\end{figure}

\emph{Spacetime close to spatial infinity.} The natural arena for a
discussion of the early radiative properties of spacetime is the
neighbourhood of $i^0$ and $\scri^+$. In particular, it has been shown
that under by \emph{fiat} assumptions that the Bondi mass tends to
the ADM mass as one approaches spatial infinity along future null
infinity \cite{AshMag79}. In order to discuss the properties of
spacetime in this region, one would like to make use of the Friedrich
conformal field equations. These provide a regular system of partial
differential equations for the field quantities in the unphysical
spacetime. However, in the way they were first given (see
for example  \cite{Fri83}) they are not too well suited for an
analysis near spatial infinity. It 
can be shown that for the Weyl tensor one has
$\psi_{abcd}=\O(m/\rho^3)$, where $\rho$ is a geodesic distance to
spatial infinity on an initial Cauchy hypersurface $S$, and $m$ is the
ADM mass.  Thus, the standard representation of spatial infinity as a
point (see figure \ref{fig:point}) is too narrow to discuss fully the
consequences of the conformal field equations.
Using a extended version \cite{Fri95} of the conformal field equations
which allows to consider Weyl connections and a gauge based on the
properties of conformal geodesics, it is possible to arrive to a new
representation of spatial infinity which allows to formulate an
initial value problem in the neighbourhood of spatial infinity such
that: (i) the equations and initial data are regular, (ii) the
location of null infinity is known a priori via a conformal factor
$\Theta$ that can be read-off from the initial data once the
constraint equations have been solved \cite{Fri98a}. The
unknowns in this setting are the frame ($c^{\mu}_{ab}$, $\mu=0,\dots,3$), the
connection ($\Gamma_{abcd}$), the Ricci curvature ($\Theta_{abcd}$)
and the Weyl curvature ($\phi_{abcd}$) of the unphysical
spacetime, $a,b,c,d=0,1$ being spinorial indices. In
this conformal geodesic gauge representation of spatial infinity, the
point $i^0$ is blown up to obtain a cylinder $I=S^2\times
(-1,1)=\{\rho=0, |\tau|<1\}$ (see
figure \ref{fig:cylinder}). The sets $I^\pm=\{\rho=0,\tau=\pm 1\}$
where null infinity touches spatial infinity are of a special nature
and will be further discussed later. 

In the gauge given by the conformal
geodesics, the equations for the unknowns
$v=(c^\mu_{ab},\Gamma_{abcd},\Theta_{abcd})$ are of the form:
\begin{equation}
\label{v_equations}
\partial_\tau v=K v+Q(v,v)+L\phi,
\end{equation}
where $K$, $Q$ denote respectively linear and quadratic functions with
constant coefficients, and $L$ denotes a linear function with
coefficients depending on the coordinates and such that
$L|_I=0$. Differentiating formally with respect to $\rho$ $p$ times and
evaluating on $I$ one obtains,
\begin{eqnarray}
\label{transport_v}
&&\hspace{-3mm}\partial_\tau v^{(p)}=Kv^{(p)}+Q(v^{(0)},v^{(p)})+Q(v^{(p)},v^{(0)}) \nonumber\\
&&
+\sum_{j=1}^{p-1}\left(Q(v^{(j)},v^{(p-j)})+L^{(j)}\phi^{(p-j)}\right)+L^{(p)}\phi^{(0)},
\end{eqnarray}
where ${}^{(p)}$ denotes the operation of taking the $p$-th derivative
with respect to $\rho$ and evaluating at $\rho=0$, and
$\phi=(\phi_0,\phi_1,\phi_2,\phi_3,\phi_4)$. $L^{(p)}\neq 0$ for
$p\geq 1$. The equations for the components of
the Weyl tensor are slightly more complicated,
\begin{equation}
\label{bianchi_equations}
A^\mu\partial_\mu \phi=B(\Gamma)\phi,
\end{equation}
where $A^\mu$, $\mu=0,\ldots,3$ are $5\times 5$ matrices depending on
$c^\mu_{ab}$ such that,
\begin{equation}
\label{matrices_at_I}
A^0|_I=\sqrt{2}\mbox{diag}(1+\tau,1,1,1,1-\tau), \qquad A^1|_I=0,
\end{equation}
and $B(\Gamma)$ is a matrix value function of the connection
coefficients. From the second equality one sees that the $\rho$
derivatives disappear from the field equations upon evaluation on $I$.
Therefore, we call $I$ a \emph{total characteristic} of the system
(\ref{v_equations})-(\ref{bianchi_equations}). Differentiating
formally with respect to $\rho$ $p$ times and evaluating on $I$ one
obtains,
\begin{eqnarray}
\label{transport_phi}
&&\hspace{-0.7cm}A^0\partial_\tau\phi^{(p)}+A^C\partial_C\phi^{(p)}=
B(\Gamma^{(0)}_{abcd})\phi^{(p)}  \nonumber \\
&&\hspace{-0.4cm}+\sum_{j=1}^p \left(\begin{array}{c} p \\ j
  \end{array}\right)\left\{B(\Gamma^{(j)}_{abcd})\phi^{(p-j)}-(A^\mu)^{(j)}\partial_\mu\phi^{(p-j)}\right\},
\end{eqnarray}
where $C=2,3$. The system composed by (\ref{transport_v}) and
(\ref{transport_phi}) for a fixed $p$ are called the \emph{transport
  equations} of order $p$ on the cylinder at spatial infinity. The
transport equations allow us to obtain expansions for
the field equations of the form,
\begin{equation}
u=\sum\frac{1}{p!} u^{(p)}\rho^p,
\end{equation}
where $u^{(p)}=u^{(p)}(\tau,x^C)$. In the case of time
symmetric initial data, the solutions of the system
(\ref{transport_v}) with $p=0$ coincide with the $p=0$ solutions of
Minkowski spacetime.  For $p\geq 1$, the equations are linear. The
equations (\ref{transport_v}) are essentially ode's which allow us to
calculate $v^{(p)}$ if $u^{(k)}$ for $0\leq k <p$ are known. Once this
has been done, one can calculate $\phi^{(p)}$. Note however, that
because of the first equality in (\ref{matrices_at_I}), the system
(\ref{transport_phi}) is singular at the sets $I^\pm=\{\rho=0,\tau=\pm
1\}$ where null infinity ``touches''spatial infinity. This phenomenon
can be understood by recalling that while $\scri$ is a characteristic
of the conformal field equations, $I$ is a total characteristic. Thus,
a degeneracy occurs at $I^\pm$.

In \cite{Fri98a} a first analysis of the properties of the transport
equations under the assumption of time symmetry of the initial data
(i.e. vanishing extrinsic curvature of the initial hypersurface) was
carried out. It is found that a necessary condition for the transport
equations to have smooth solutions at the sets $I^\pm$ is,
\begin{equation}
\label{regcond}
D_{(a_sb_s\cdots}D_{a_1b_1}b_{abcd)}(i^0)=0, \quad s=0,1,\dots \;\; , 
\end{equation}
where $D_{ab}$ denotes covariant derivatives with respect to the
initial 3-metric, and $b_{abcd}$ is the Cotton-Bach spinor. The
vanishing of the 
Cotton-Bach spinor characterises locally conformally flat initial
data. If the regularity condition (\ref{regcond}) does not hold for a
certain $p$ it can be seen that the transport equations develop
logarithmic singularities at $I^\pm$. It is conjectured that the
regularity condition is also a sufficient condition for the solutions of the
transport equations with time symmetric initial data to be smooth at
$I^\pm$. This however remains to be proved. The generalisation of the
condition (\ref{regcond}) for non-time symmetric is also unknown. The
importance of the transport equations (\ref{transport_v}) and
(\ref{transport_phi}) lies in the fact that they allow us to relate
directly quantities defined on $\scri$ ---the Bondi mass in
particular--- with the initial data from which the asymptotically
simple spacetime arises. 

\emph{Conformally flat initial data.} The regularity condition
(\ref{regcond}) is satisfied in a trivial way for time symmetric,
conformally flat initial data. This fact allow us to conjecture that
the developments of this class of initial data are candidates for
spacetimes with a smooth null infinity. Conformally flat initial data
are of great interest as all the time symmetric initial data sets for
head-on black hole collisions currently in use in numerical
simulations are of this type. We will restrict the following
discussion to them.  Being time symmetric, the initial 3-geometries we
want to consider are completely described in terms of the following
line element,
\begin{equation}
ds^2=-\chi^4\;ds^2_{flat},
\end{equation}
with $\chi$ satisfying the Laplace equation. It is customary to write,
$\chi=1+\rho W$, where $\rho$ is the geodesic distance from $i$. From
the Yamabe equation one has that,
\begin{equation}
\label{W}
W=\frac{m}{2}+\sum_{p=1}^\infty \sum_{k=-p}^p \frac{1}{p!}w_{p,2p,n-k}i^{2n-m}\sqrt{\frac{4\pi}{2n+1}}Y_{n,m}\rho^p,
\end{equation}
where $m$ is the total (ADM) mass, and $w_{p,2p,n-k}$ are complex
constants satisfying the reality condition
$\overline{w}_{p,2p,2p-k}=w_{p,2p,k}$. The functions $Y_{n,m}$ are the
standard spherical harmonics. Although the system of transport
equations is of large size ($50\times 50$), its properties make it quite amenable to a treatment by means
of a computer algebra system. A series of scripts in the computer
algebra system Maple V have been written in order to solve the
transport equations (\ref{transport_v}) and (\ref{transport_phi}) up
to $p=5$ for time symmetric conformally flat initial data. The
solutions of the transport equations up to this order are polynomial.
This is in itself an outstanding feature, and it is believed to occur
at all higher orders. Yet a proof of this fact lies far in the future.
This adds further support to the conjecture that time symmetric,
conformally flat initial data sets yield developments with a smooth
null infinity. The details of the implementation of the transport
equation solver and the detailed structure of the solutions for
generic time symmetric initial data 
will be presented elsewhere. It should be pointed out that the
transport equations are extremely sensitive to any small change in
their structure. Any change would inmediately give rise to non-smooth
solutions containing logarithms. The absense of such kind of solutions
in the expansions up to order $p=5$ is in itself a guarantee of the
correctness of our results.

\emph{Calculating the Bondi mass near spatial infinity.} A natural
application of the expansions is to calculate the first orders of the
Bondi mass in aregion of $\scri$ close to spatial infinity. The
formula (\ref{bondi_mass}) cannot be used directly as it stands for it
is given in the so-called Bondi gauge, while the gauge used in the
representation of spatial infinity is based in the properties of
conformal geodesics. The connection between the two gauges has been
given in \cite{FriKan00}. Knowledge of the solutions to the transport
equations up to $p=5$ order permits the calculation of the first 2
terms of the expansion for the Bondi mass. A lengthy calculation
yields,
\begin{equation}
\label{bondi_mass:expansion}
m_B=m+\sum_{k=-2}^2 |G_k|^2\left(\frac{\sqrt{2}}{u}\right)^7 +O(\rho^8),
\end{equation}
where $u$ is a standard Bondi retarded time ($u\rightarrow -\infty$ as
one approaches to $I^+$) and,
\begin{eqnarray*}
&&G_{-2}=-\textstyle\frac{\sqrt{6}}{4}\left(mw_{2;4,4}-2\sqrt{6}w_{1;2,2}^2\right), \\
&&G_{-1}=-\textstyle\frac{\sqrt{6}}{4}\left(mw_{2;4,3}-2\sqrt{12}w_{1;2,2}w_{1;2,1}\right), \\
&&G_{0}=-\textstyle\frac{\sqrt{6}}{4}\left(mw_{2;4,2}-4(w_{1;2,0}w_{1;2,2}+w_{1;2,1}^2)\right), \\
&&G_{1}=-\textstyle\frac{\sqrt{6}}{4}\left(mw_{2;4,1}-2\sqrt{12}w_{1;2,0}w_{1;2,1}\right), \\
&&G_{2}=-\textstyle\frac{\sqrt{6}}{4}\left(mw_{2;4,0}-2\sqrt{6}w_{1;2,0}^2\right),
\end{eqnarray*}
are, up to an irrelevant numerical factor, the \emph{Newman-Penrose
  (NP) constants} \cite{NewPen65} of the development expressed in
terms of the initial data as given by Friedrich and K\'{a}nn\'{a}r.
Note that close to $I^+$ the second term in
(\ref{bondi_mass:expansion}) is negative, in agreement with the fact
that the Bondi mass is non-increasing. The formula
(\ref{bondi_mass:expansion}) is rather unexpected. Before having
performed the calculation, there was principle, no reason for this to
be the case. This evidenciates the power conveyed in our approach, and
the kind of information that it is possible to extract through it. The
NP constants are defined as integrals over cuts of $\scri^+$ of the
derivative with respect of a parameter of the generators of light
cones escaping to null infinity of the $\psi_0$ component of the Weyl
tensor. They are absolutely conserved in the sense that they have the
same value for any cut of $\scri^+$. For spacetimes admitting a
conformal compactification that includes the point $i^+$ (future time
infinity), the NP constants are equal to the value of the Weyl tensor
precisely at $i^+$ \cite{NewPen68,FriSch87}.  Thus, the NP constants
contain global information of the spacetime that can be read directly
from the initial data! The NP constants have also been found playing a
role in determining the decay rate of self-gravitating waves in a
Schwarzschild background \cite{GomWinSch94}:waves with vanishing NP
constants are found to decay faster. On the same lines, an analysis of
some particular boost-rotation symmetric solutions near $i^+$ has
shown, remarkably, that the squared norm of the NP constants appear
again in the leading term of the time derivative of the Bondi mass
(late radiated power) \cite{LazVal00}. There has been some discussion
concerning the idea of the NP giving a measure of the spurious
radiation contained in the spacetime in general, and via formula
(\ref{bondi_mass:expansion}) on the initial Cauchy hypersurface in
particular. The idea being that the first radiation reaching $\scri^+$
as the system begins to evolve can not be truthly attributed to the
processes occurring in the interior of the spacetime (a black hole
collision, for example) as it requires some time to reach the
asymptotic region. Formula (\ref{bondi_mass:expansion}) is local, and
thus valid for an arbitrarily small neighbourhood of spatial infinity.
It will register radiation that is already in the asymptotic region
and that had to be put in in order to construct the initial data. Note
that formula (\ref{bondi_mass:expansion}) implies that the NP
constants of static spacetimes (the Schwarzschild solution in
particular) arising from conformally flat data must vanish. This
latter fact can also be independently verified using the
Friedrich-K\'{a}nn\'{a}r formula for the NP constants on static
initial data. Similarly, the recent examples of non-trivial
asymptotically simple spacetimes given by Chru\'{s}ciel and Delay
\cite{ChrDel02} have also vanishing NP constants, and accordingly a
very special radiation content.

The formula (\ref{bondi_mass:expansion}) has only been calculated for
developments of time symmetric, conformally flat initial data sets. It
is expected that a similar result would hold (modulo some more
involved calculations) for general time symmetric data. For initial
data with non-vanishing second fundamental form the situation is less
clear. The study expansions of the gravitational field obtained from
the transport equations (\ref{transport_v}) and (\ref{transport_phi})
were first carried with the purpose of laying the first steps towards
a global existence proof of non-trivial spacetimes with a smooth
structure at null infinity in general, and the verification of the
conjecture regarding the regularity condition (\ref{regcond}) in
particular. However, as (\ref{bondi_mass:expansion}) evidenciates they
also happen to be a powerful tool in order to discuss the physics
contained in these spacetimes.

\emph{Comparing the Misner and Brill-Lindquist initial data sets.} As
an application of our result, one can perform a comparison of the
NP constants of the development of physically equivalent
Brill-Lindquist (BL) and Misner initial data. The interest of this example
lies in the fact that both initial data sets provide a model to
describe head on black hole collisions. Both initial data sets are
time symmetric, conformally flat, and the crucial difference between
each other lies in their topology: 3 asymptotically flat regions
connected by 2 throats in the BL case, and 2 asymptotically
flat regions connected by 2 throats in the Misner case. For physically
equivalent BL and Misner data sets it is understood data
sets such that: (1) the ADM mass of the initial hypersurface measured
at the reference asymptotic end is the same; and (2) the geodesic
separation of the innermost minimal surfaces is the same. For the case
of the Misner data, there is an explicit formula for the separation of
the minimal surfaces, while for the BL data it has to be
calculated numerically. \v{C}ade\v{z} \cite{Cad74} has given a
tabulation of the values of this distance for different values of the
parameters of the BL initial data.
\begin{figure}[t]
\includegraphics[width=.30\textwidth]{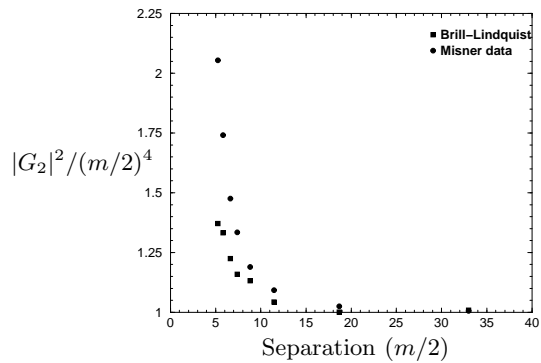}
\put(-105,-10){Separation $(m/2)$}
\put(-200,60){$|G_2|^2/(m/2)^4$}  
\caption{\label{fig:comparison} The Newman-Penrose constants for
  physically equivalent Misner and Brill-Lindquist initial data sets.}
\end{figure}
Due to the axial symmetry of the Misner and BL
initial data sets, there is only one non-vanishing NP
constant for each one of the developments ($G_2$). An analysis of the
NP constants for the BL data has been recently given in
\cite{DaiVal02}. The function $W$ ---see formula (\ref{W})--- for the
BL data has been calculated there. The one for the Misner data can be
calculated in a similar fashion. A plot of the
values of the constant for various physically equivalent initial data
sets is shown in figure \ref{fig:comparison}. It can be seen that for black
holes initially very separated the difference between the
NP constants of the two topologically different initial
data set is negligible, and it
increases as one considers data sets where the separation is
smaller. For black holes separated a distance of $L\approx 5.26 (m/2)$
where $m$ is the total ADM mass of the spacetime one has $|G_{2
  (BL)}|^2/|G_{2_(Misner)}|^2\approx 0.66$. The physical content of
this difference, and the possibility of interpreting it as some
measure of the (spurious) radiation contained in the initial data sets
is certainly a matter of debate.

\emph{Acknowledgements.} I would like to thank H. Friedrich, S. Dain,
J. Winicour and M. Mars for several discussions.


\end{document}